\documentclass[%
reprint,pra,
superscriptaddress,
amsmath,amssymb,
aps,
]{revtex4}

\usepackage{graphicx}
\usepackage{dcolumn}
\usepackage{bm}
\usepackage{hyperref}
\usepackage[utf8]{inputenc}
\usepackage{mathrsfs}
\usepackage{subfigure}
\usepackage{cleveref}
\usepackage{amsmath}
\usepackage[T1]{fontenc}
\usepackage{mathptmx}
\usepackage{multirow}
\draft 

\begin{document}
	
	\preprint{APS/123-QED}

\title{High order Coherence Functions and Spectral Distributions as given by the Quantum Theory of Laser Radiation}

\author{Tao Peng}%
\affiliation{%
   Texas A\&M University, College Station, Texas, 77843, USA}%
\author{Xingchen Zhao}%
\affiliation{%
	Texas A\&M University, College Station, Texas, 77843, USA}%
\author{Yanhua Shih}%
\affiliation{%
	University of Maryland, Baltimore County, Baltimore, Maryland 21250, USA}%
\author{Marlan O. Scully}%
\email{scully@tamu.edu}
\affiliation{%
	Texas A\&M University, College Station, Texas, 77843, USA}%
\affiliation{%
	Baylor University, Waco, 76706, USA}%
\affiliation{%
	Princeton University, Princeton, New Jersey 08544, USA}%

\begin{abstract}

We propose and demonstrate a method for measuring the time evolution of the off-diagonal elements $\rho_{n,n+k}(t)$ of the reduced density matrix obtained from the quantum theory of the laser. The decay rates of the off-diagonal matrix element $\rho_{n,n+k}(t)$ (k=2,3) are measured for the first time and compared with that of $\rho_{n,n+1}(t)$, which corresponds to the linewidth of the laser. The experimental results agree with the quantum theory of the laser.

\end{abstract}

\maketitle
\section{Introduction}

Quantum coherence effects in molecular physics are largely based on the existence of the laser \cite{pestov2007optimizing}. Indeed, in most of our experiments and calculations we take the laser to be an ideal monochromatic light source. If the laser linewidth is important then we usually just include a ``phase diffusion'' linewidth into the logic. But what if we are thinking about higher order correlation effects in an ensemble of coherently driven molecules. For example, photon correlation and light beating spectroscopy involving Glauber second order correlation functions \cite{cummins2013photon, glauber1963quantum}. Furthermore, third and higher order photon correlations of the laser used to drive our molecular system can be important. The investigation of higher order quantum laser noise is the focus of the present paper. 

\begin{figure}[hbt!]
	\begin{center}
		\includegraphics[width=10 cm]{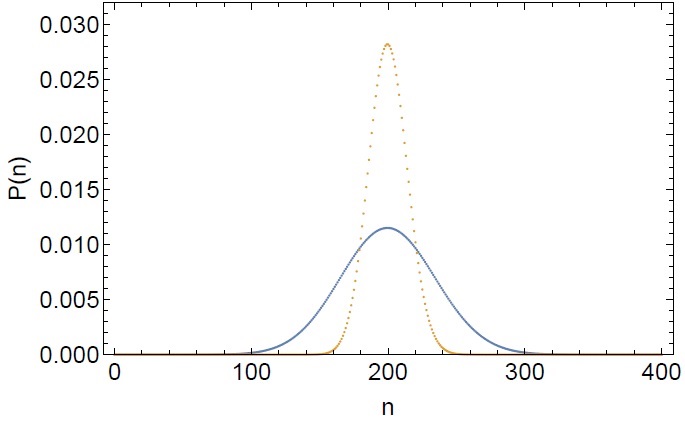}
	\end{center}
	\caption{Steady state photon distribution function for coherent (orange dashed line) and laser radiation (blue solid line). The laser is taken to be 20 percent above threshold, $\langle n \rangle =200$.}\label{fig:comparison}
\end{figure}

Fifty years ago the quantum theory of the laser (QTL) was developed using a density matrix formalism \cite{scully1966quantum}. In the interesting threshold region \cite{haken1966theory,degiorgio1970analogy} the steady state laser photon statistics is given by the diagonal elements of the laser density matrix as
\begin{align}\label{rho}
\rho_{n,n} =\mathfrak{N} \prod\limits^{n}_{m=0}[\alpha-\beta m]/\gamma,
\end{align}
where $\alpha$ is the linear gain, $\beta$ is the nonlinear saturation coefficient, $\gamma$ is the cavity loss rate, and $\mathfrak{N}$ is the normalization constant:
\begin{align}\label{normalization_constant}
\mathfrak{N}^{-1} =\sum\limits_n{\prod\limits^{n}_{m=0}[\alpha-\beta m]/\gamma}.
\end{align}
Eq.~(\ref{rho}) is plotted in Fig.~\ref{fig:comparison} where it is compared with a coherent state. 

The formalism developed in the QTL density matrix analysis has since been successfully applied to many other physical systems such as the single-atom maser(aka the micromaser) \cite{meschede1985one}, the Bose-Einstein condensate (aka the atom laser, see Table \ref{table:parameters}) \cite{scully1999condensation}, pion physics \cite{hoang1997remarks}, etc. Other applications of the formalism have been developed recently and more will likely emerge. Thus we are motivated to deeper our understanding of the QTL by further analyzing and experimentally verifying the time dependence of off-diagonal elements $\rho_{n,n+k}(t)\equiv\rho^{(k)}_{n}(t)$. The diagonal elements of the laser density matrix for which $k=0$, have been well studied. Not as for the off-diagonal elements. In particular $\rho^{(1)}_n (t)$ yields the Schawlow-Townes laser linewidth. But what about the higher order correlations $k=2,3 \cdots$? That is the focus of the current paper.

\section{Theory and Experiment}
\begin{figure}[hbt!]
	\begin{center}
		\includegraphics[width=10 cm]{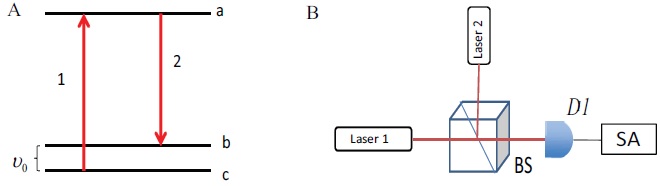}
	\end{center}
	\caption{(A) Conceptual illustration of laser spectral measurement by beating two laser fields laser 1 and 2. The frequency difference between the two fields is $\nu_0$. The lower level doublet may be thought as hyperfine doublet whose dipole is driven at frequency $\nu_0$ and detected by a pick-up coil (not shown) in the usual way. (B) Experimental setup used in measuring the spectrum of the beat note between the laser and local oscillator. The beat note signal is measured by the detector ($D1$) and analyzed by the spectrum analyzer(SA). BS, non-polarizing beamsplitter.}\label{fig:setup_1}
\end{figure}

The off-diagonal elements vanish at steady state, regressing to zero as \cite{scully1966quantum}
\begin{align}\label{phase}
\rho^{(k)}_{n}(t)=\rho^{(k)}_{n}(0)\text{exp}(-k^2D t)
\end{align}
where  $D=\gamma/\bar{n}$ is the Schawlow-Townes  phase diffusion linewidth and $\bar{n}=(\alpha-\gamma)/\beta$.
The expectation value of the laser amplitude operator is given by
\begin{align}\label{E_rho}
\langle \hat{E}(z,t)\rangle=\mathscr{E}_{0}\sin\kappa z \sum_{n}{\rho^{(1)}_{n}(0)\sqrt{n+1}e^{-Dt}e^{i\nu t}},
\end{align}
where $\nu$ is the center frequency of the laser field and the electric field per photon is given by $\mathscr{E}_0=\sqrt{\hbar\nu/\epsilon_{0}V}$, where $\epsilon_{0}$ is the permittivity of free space and V is the laser cavity volume. The physics is explained in Fig.~\ref{fig:setup_1} and associated text.

\begin{figure}
	\begin{center}
		\includegraphics[width=10 cm]{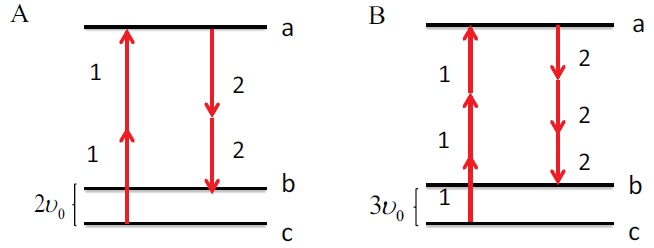}
	\end{center}
	\caption{Conceptual illustration of the second order (A) and third-order (B) correlated spectral measurement. Again, the frequency difference between the two laser fields is $\nu_0$. The lower level doublet may be thought as hyperfine doublet (triplet) whose dipole is driven at frequency $2\nu_0$ ($3\nu_0$) and detected by a pick-up coil.}\label{fig:setup_2}
\end{figure}

As is discussed in the following, the second order off-diagonal elements are given by the field operator averages 
\begin{align}\label{E-square-rho}
&\ \ \ \ \langle \hat{E}(z,t)\hat{E}(z,t)\rangle \cr
&=\mathscr{E}^2_{0}\sin^2\kappa z\sum_n{\rho^{(2)}_{n}(0)\sqrt{(n+1)(n+2)}e^{-4Dt}e^{i2\nu t}},
\end{align}
and the third order off-diagonal elements are given by
\begin{align}
&\ \ \ \ \langle \hat{E}(z,t)\hat{E}(z,t)\hat{E}(z,t)\rangle \cr
&=\mathscr{E}^3_{0}\sin^3\kappa z\small{\sum_n}{\rho^{(3)}_{n}(0)\sqrt{(n+1)(n+2)(n+3)}e^{-9Dt}e^{i3\nu t}},\cr 
&\label{E-triple-rho}
\end{align}
where again the physics is explained in Fig.~\ref{fig:setup_2} and associated text.

Eq.~(\ref{E_rho}) gives the time evolution associated with the first order off-diagonal elements $\rho^{(1)}_n$, yielding the spectral profile of the laser. The heterodyne method is usually adapted to measure the linewidth of the laser \cite{okoshi1980novel,richter1986linewidth}, in which case the center frequency is shifted from optical frequency to the radio frequency range. A natural way to measure the laser linewidth is to beat two almost identical but uncorrelated lasers \cite{muanzuala2015measuring} such that the beat frequency between the lasers is in the MHz range. The result, as seen from Eq.~(\ref{spectral_1}), is twice of the laser linewidth when the two independent lasers are nearly identical. 

Many experiments have been carried out to determine the linewidth \cite{okoshi1980novel} and photon statistics \cite{arecchi1965measurement} of the laser. Other experiments have measured the intensity correlation of the laser at threshold \cite{corti1973measurements}, revealing the influence of the intensity fluctuation on the laser spectrum. However, to the best of our knowledge, no measurements have been made of the higher order phase correlations ($k\geq 2$).  Here we measure the second and third correlation of the heterodyne signals from two independent lasers, which yields the second  and third order time evolution of a laser above threshold. Specifically, we performed the following experiments: the first set of experiments is to measure the spectral profile of the laser beat note, i.e., allows us to measure the decay rate as shown in Eq.~(\ref{E_rho}). The other two sets of experiments determine the spectral profile of the second and third order correlated beat notes, this allows us to measure the decay rate as shown in Eq.~(\ref{E-square-rho}) and Eq.~(\ref{E-triple-rho}).
\begin{table*}[!ht]
	\begin{tabular}{ |l|c|c| }
		\hline
		& Laser & BEC \\ \hline
		\multirow{3}{*}{\quad $\alpha$ \quad}
		& {} & {} \\
		& {Linear stimulated emission gain} & {Rate of cooling due to interaction} \\
		& {} & {with walls times the number of atom N} \\
		\hline
		\multirow{3}{*}{\quad $\beta$\quad}
		& {} & {} \\
		& {Nonlinear saturation due to the reabsorption} & {Nonlinearity parameter due to the constraint that} \\
		& {of photons generated by stimulated emission} & {there are N atoms in the BEC: } \\ &{} &{numerically equal to $\alpha/N$.}\\
		\hline
		\multirow{3}{*}{\quad$\gamma$\quad}
		& {} & {} \\
		& {Loss rate due to photons absorbed} & {Loss rate due to photon absorption from the} \\
		& {in cavity mirrors etc.} & {thermal bath (walls) equal to $\alpha(T/T_c)^3$.} \\ 
		\hline
	\end{tabular}
	\caption{Parameters in laser and BEC systems.}
	\label{table:parameters}
\end{table*}
The laser spectral measurement is conceptually illustrated in Fig.~\ref{fig:setup_1}A. The frequency difference between the two laser fields of laser 1 and 2 is $\nu_0\equiv\nu_1 -\nu_2$, $\nu_1$ and $\nu_2$ represent the center frequencies of the laser 1 and 2, respectively. The lower level doublet may be thought as hyperfine doublet whose dipole is driven at frequency $\nu_0$ and detected by a pick-up coil. Fig.~\ref{fig:setup_1}B illustrates the setup of the first set of experiments. This is a typical heterodyne detection setup, the center frequency between the two He-Ne lasers is in the MHz range. This difference allows us to analyze the beat signal around a non-zero value hence the full shape of the linewidth is obtained unambiguously. A non-polarizing beamsplitter (BS) is used to mix the two laser beams. The beat signal is then directed to the photodiode ($D1$) after the BS. A fast Fourier transform (FFT) of the signal is performed by the spectrum analyzer (SA) giving the frequency spectrum of the beat note.   

\begin{figure}[hbt!]
	\begin{center}
		\includegraphics[width=0.4\linewidth]{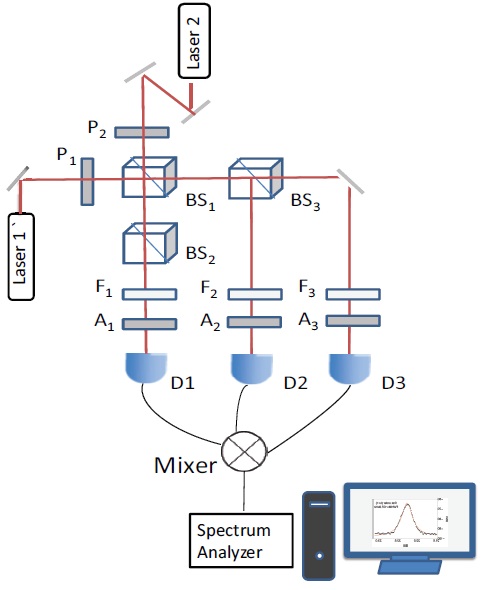}
	\end{center}	
	\caption{Schematic setup for measuring higher order spectral line distribution up to 3rd order. Laser 1 and 2 : He-Ne lasers; P: polarizer; F: filter; A: analyzer; BS: non-polarizing beamsplitter; Mixer: frequency mixer; $D1$, $D2$, and $D3$, photodiode detectors.}\label{fig:setup_3}
\end{figure}

For the first set of experiments, the first order coherence function \cite{glauber1963quantum,scully1966quantum} is 
\begin{align}\label{G1_1}
G^{(1)}(t)&=\mathrm{Tr}\{\rho[(\hat{E}^{\dagger}_{1}(t)+\hat{E}^{\dagger}_{2}(t))(\hat{E}_{1}(t)+\hat{E}_{2}(t))]\}\cr
&=\mathrm{Tr}\{(\rho_1\otimes\rho_2) [|\hat{E}_{1}(t)|^2+|\hat{E}_{2}(t)|^2+\hat{E}^{\dagger}_{1}(t)\hat{E}_{2}(t)+c.c.\}\cr
&=\mathscr{E}^2_1\mathrm{Tr}[\rho_1\hat{a}^{\dagger}_{1}(t)\hat{a}_{1}(t)]+\mathscr{E}^2_2\mathrm{Tr}[\rho_2\hat{a}^{\dagger}_{2}(t)\hat{a}_{2}(t)]\cr&+\mathscr{E}_1\mathscr{E}_2\{\mathrm{Tr}[(\rho_1\otimes\rho_2)\hat{a}^{\dagger}_{1}(t)\hat{a}_{2}(t)]e^{i(\nu_1 -\nu_2)t}+c.c.]\},
\end{align}
where $\rho=\rho_1\otimes\rho_2$ is the density operator of the system, $\rho_1$ and $\rho_2$ represent the density operators of laser 1 and 2,  respectively. 
From the above equation, we can see the only terms carry the beat note frequency are 
\begin{align}\label{gamma_nu}
\Gamma^{(1)}(t)=\mathscr{E}_1\mathscr{E}_2\mathrm{Tr}[(\rho_1\otimes\rho_2)\hat{a}^{\dagger}_{1}(t)\hat{a}_{2}(t)]e^{i\nu_0 t},
\end{align}
with its complex conjugate which contributes to the $-\nu_0$ frequency component. Under the condition that the two lasers are independent, we can rewrite Eq.(\ref{gamma_nu}) as
\begin{align}\label{first_coherence}
\Gamma^{(1)}(\nu_0, t) =\mathscr{E}_{1}\sum_{n_1}{\sqrt{n_1 +1}\rho^{(1)}_{n_1}(0)e^{-D_{1} t}}\times \mathscr{E}_{LO}\sum_{n_{2}}{\sqrt{n_{2}}\rho^{(-1)}_{n_{2}}(0)e^{-D_{2} t}}e^{i\nu_0 t}.
\end{align}
Taking the Fourier transform, we have a Lorentzian spectrum centered at the beat frequency $\nu_0$ with a width $D'=D_1 +D_{2}$, which is essentially twice the width of one laser
\begin{align}\label{spectral_1}
S_{\nu_0}(\omega)\propto\frac{D'}{(\omega -\nu_0)^2+(D')^2}.
\end{align}
As shown in Fig.~\ref{fig:setup_2}, the higher order spectral measurements can be conceptually understood in a similar way as the laser linewidth measurement. The frequency difference between the two laser fields is again $\nu_0$. The lower level doublet can be thought as hyperfine doublet whose dipole is driven at frequency $2\nu_0$ or $3\nu_0$, respectively, and detected by a pick-up coil. 
The second and third experiments measure the spectral profile of the second  and third order correlation of beat notes, the setup is shown in Fig.~\ref{fig:setup_3}. We used the same two lasers to create the beat signal, where three detectors $Di (i=1,2,3)$ are used. The outputs from the photodiodes are used as inputs for a frequency mixer. The output from the mixer is then sent to the spectrum analyzer and the frequency spectrum of the correlated signal is obtained after the FFT.
As shown in Fig.~\ref{fig:setup_3}, this set of experiments measures the laser field correlation that is governed by the time evolution of the second  and third order off-diagonal elements $\rho^{(2)}_n (t)$ and $\rho^{(3)}_n (t)$, respectively. The quantity we now measure is determined by the correlation of the heterodyne signals from detectors as in Fig.~\ref{fig:setup_3}. We have the signal of interest at frequency $2\nu_0$ from the second order  coherence function is
\begin{align}\label{2nu}
\Gamma^{(2)}(t)=\mathscr{E}^2_1\mathscr{E}^2_2\mathrm{Tr}(\rho_1\otimes\rho_2)\hat{a}^{\dagger}_{1}(t)\hat{a}^{\dagger}_{1}(t)\hat{a}_{2}(t)\hat{a}_{2}(t)e^{i2\nu_0 t},
\end{align}
with its complex conjugate contributes to the $-2\nu_0$ frequency. The correlated heterodyne signal is
\begin{align}\label{second_coherence}
\Gamma^{(2)}(2 \nu_0, t)&=\mathscr{E}^2_{1}\sum_{n_1}{\rho^{(2)}_{n_1}(0)\sqrt{(n_1 +2)(n_1 +1)}e^{-4D_1 t}}\cr
&\times\mathscr{E}^2_{2}\sum_{n_{2}}{\rho^{(-2)}_{n_{2}}(0)\sqrt{(n_{2}-1)n_2}e^{-4D_{2} t}}e^{i2\nu_0 t}.
\end{align}
Taking the Fourier transform, we get a Lorentzian spectral profile centered at $2\nu_0$ with a width of $4D'$
\begin{align}\label{spectral_2}
S_{2\nu_0}(\omega)\propto\frac{4D'}{(\omega -2\nu_0)^2+(4D')^2}.
\end{align}
similarly, we have 
\begin{align}\label{spectral_3}
S_{3\nu_0}(\omega)\propto\frac{9D'}{(\omega -3\nu_0)^2+(9D')^2}.
\end{align}

\begin{figure}[hbt!]
	\begin{center}
		\includegraphics[width=\linewidth]{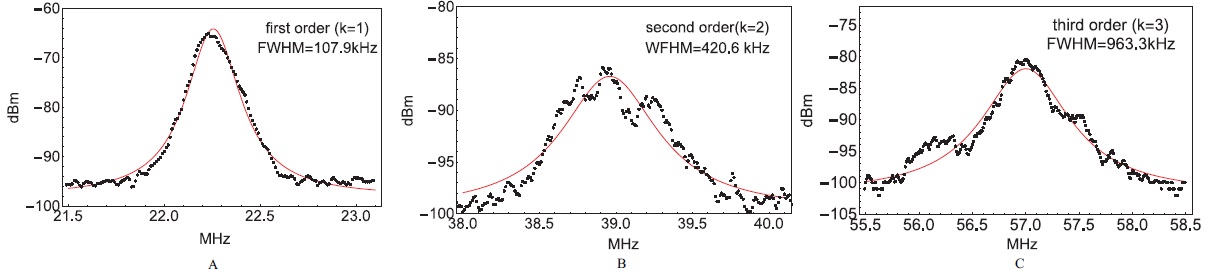}
	\end{center}
	\caption{Experimental results from the two sets of measurements. The bandwidths of the detectors are 50 MHz, the resolution bandwidth of the SA is 10 kHZ. The black dots are experimental data and the red curves are theory. (A) is the beat signals from $D1$, where the FWHM is 107.9 kHz with average 50 times. Theory is the Fourier transform of the laser fields time evolution ($e^{-D't}$) associated with frequency $\nu_0$, as shown in Eq.~(\ref{spectral_1}); (B) is correlated signal from $D1$ and $D2$, where the FWHM bandwidth is 420.6 kHz with average 50 times. Theory is the Fourier transform of the correlated laser fields time evolution ($e^{-4D't}$) associated with frequency $2\nu_0$, as shown in Eq.~(\ref{spectral_2}).(C) is correlated signal from $D1$, $D2$, and $D3$, where the FWHM is 963.3 kHz with average 50 times. Theory is the Fourier transform of the correlated laser fields time evolution ($e^{-9D't}$) associated with frequency $3\nu_0$, as shown in Eq.~(\ref{spectral_3}).}\label{fig:measurements}
\end{figure}

The main experimental results are shown in Fig.~\ref{fig:measurements}. All measurements were taken with the laser operating at the same average output power level. The resolution bandwidth (RBW) of the SA is 10 kHz, video bandwidth (VBW) is 30 kHz in all the measurements. For the sake of simplicity, the Full width at half maximum (FWHM) linewidth is taken at the -3 dB width of the measured spectrum by considering only the Lorentzian fitting \cite{muanzuala2015measuring}.

Fig.~\ref{fig:measurements}A represents the data of the first set of experiments with an average of 50 measurements of beat note signal from $D1$. The theoretical fitting in the red solid line is based on Eq.~(\ref{spectral_1}), and the FWHM is 107.9 kHz. Fig.~\ref{fig:measurements}B represents the data of the second set of experiments with 50 measurements of correlated beat note signals from $D1$ and $D2$. The theoretical fitting in the red solid line is based on Eq.~(\ref{spectral_2}), and the FWHM is estimated to be 420.6 kHz. Fig.~\ref{fig:measurements}C represents the data of the third order experiments with 50 measurements from all three detectors. The theoretical fitting in the red solid line is based on Eq.~(\ref{spectral_3}), and the FWHM is estimated to be 963.3 kHz,. First of all, we see that the obtained linewidth from the second order correlation spectrum is essentially 4 times wider than that of the single beat note linewidth, as well as the third order spectrum is 9 times wider than that of the single beat note linewidth, validating our theoretical expectation. Secondly, we see that the theoretical curves fit the data well in the center peak, but not as good at the tails. This is mainly due to the influences from other noises that also contribute to the spectral profile. For the same reason, we see that the single beat note signal can be better fitted than the second  and third order correlation signals. This is mainly due to our remeasured higher order spectral signal is close to the noise level of the detection system, further using a more intense local oscillator and sensitive detection system (detector and spectral analyzer) should be able to solve this issue. Nevertheless, our data confirms the Lorentizan spectral profile of the signal and the time evolution described by Eq.~(\ref{phase}), in the case of $k=1$, $k=2$, and $k=3$.

\section{Conclusion}
In conclusion, we have studied the time evolution in the laser. We particularly measured the bandwidth of the laser beat note and the bandwidth of the correlated laser beat note, which reveal the evolution of the first, second, and third order off-diagonal elements of the laser density operator. The higher order spectra reveal the influence of the randomness in the phase of the laser field due to quantum fluctuation. Experimental results agreed with the QTL showing that the bandwidth of the third order and second order spectral profile are nine times and four times wider than that of the first order spectral profile, respectively. 

\section*{Conflict of Interest Statement}

The authors declare that the research was conducted in the absence of any commercial or financial relationships that could be construed as a potential conflict of interest.

\section*{Author Contributions}

TP, YS and MOS discussed the design priciple. TP YS and MOS derived the theory. TP and XZ performed the experiment and data analysis. TP and MOS wrote the paper. All the authors commented on the paper. 

\section*{Funding}
Air Force Office of Scientific Research (Award No. FA9550-20-1-0366 DEF), Office of Naval Research (Award No. N00014-20-1-2184), Robert A. Welch Foundation (Grant No. A-1261), National Science Foundation (Grant No. PHY-2013771), King Abdulaziz City for Science and Technology (KACST).

\section*{Acknowledgments}
We thank Z. H. Yi, R. Nessler, H. Cai, and J. Sprigg for helpful discussion.

\end{document}